\documentclass{aa}
\usepackage{graphics}
\def\comment#1{{}}

\thesaurus{02    
  (02.07.1;      
   08.16.6;      
   08.16.7 Crab; 
   13.07.2)}     

\begin{document}

\title{Pulsar Radiation and Quantum Gravity}

\author{Philip Kaaret}

\institute{Harvard-Smithsonian Center for Astrophysics,\\ 
60 Garden Street, Cambridge, MA  02138 USA\\ 
email: pkaaret@cfa.harvard.edu}

\date{Received ; accepted }

\maketitle

\begin{abstract}

Quantum gravity may lead to an energy dependence in the speed
of light.  The high energy radiation from gamma-ray pulsars
can be used to place limits on such effects.  We find that
emission from the Crab pulsar at energies above 2~GeV trails
that at 70--100~MeV by no more than 0.35~ms (95\% confidence)
and place a lower bound on the energy scale of quantum
gravitational effects on the speed of light of  $1.8 \times
10^{15} \rm \, GeV$.  This bound might be improved by two
orders of magnitude by observation of pulsations from the Crab
at higher energies, 50--100~GeV, in the near future.


\end{abstract}

\section{Introduction}

Quantum gravity may cause modification of the dispersion
relation for photons at high energies.  It has recently been
suggested that certain quantum gravity models may lead to a
first order correction to the dispersion relation which can
be parameterized as $\Delta t = L \Delta E / c E_{QG}$, where
$\Delta t$ is the magnitude of the travel time difference
between two photons whose energies differ by $\Delta E$ and
that have traveled a distance $L$, and $E_{QG}$ is the energy
scale of the dispersion effects (\cite{amelino}).  To probe
dispersion effects at high energy scales, accurate relative
timing of nearly simultaneously produced photons of different
energies which have traveled long distances is required.  Use
of sub-millisecond time structure of the keV photon flux of
gamma-ray bursts at cosmological distances (\cite{amelino};
\cite{schaefer}) and use of several minute time structure in
TeV flares from AGN (\cite{biller}) have been suggested. 
Here, we show that sub-millisecond timing of GeV emission
from gamma-ray pulsars may also place useful constraints on
the dispersion relation for photons at high energies.

Below, we use existing gamma-ray data to determine the
accuracy with which high energy pulsar emission can be timed
and to place bounds on the energy scale for quantum gravity
corrections to the speed of light.  We then discuss how this
limit might be improved by pulsar observations at higher
energies in the near future.

\section{Gamma-ray pulsations from the Crab}

We chose to analyze the energy dependence of pulse arrival
times from the Crab pulsar as it has the largest ratio of
distance to pulse period of the bright gamma-ray pulsars, thus
maximizing the constraints which can be placed.  Also, the
pulses from the Crab are well aligned in time from radio
waves, through optical and x-ray emission, to gamma-rays. 
Thus, it is likely that the photons of different energies are
produced nearly simultaneously.

We used data from the {\it Energetic Gamma-Ray Experiment
Telescope} (EGRET) (\cite{thompson}) of the {\it Compton
Gamma-Ray Observatory} (CGRO).  We extracted gamma-ray photon
event lists from the CGRO public archive for observations
pointed within $40^{\circ}$ of the Crab and then, using the
program pulsar (version 3.2, available from the CGRO Science
Support Center) selected events lying within the energy
dependent 68\% point spread function of EGRET and calculated
the phase of each photon relative to the radio ephemeris of
the Crab (\cite{arzoumanian}). The pulse period of the Crab
changed from 33.39~ms to 33.49~ms over the course of these
observations.  The radio timing must be corrected for the
variable dispersion along the line of sight to the Crab.  The
accuracy of this correction is estimated to be 0.2~ms
(\cite{nice}), consistent with previous estimates of the
accuracy of the dispersion correction for the Crab
(\cite{gullahorn}).

\begin{figure}
\resizebox{\hsize}{!}{\includegraphics{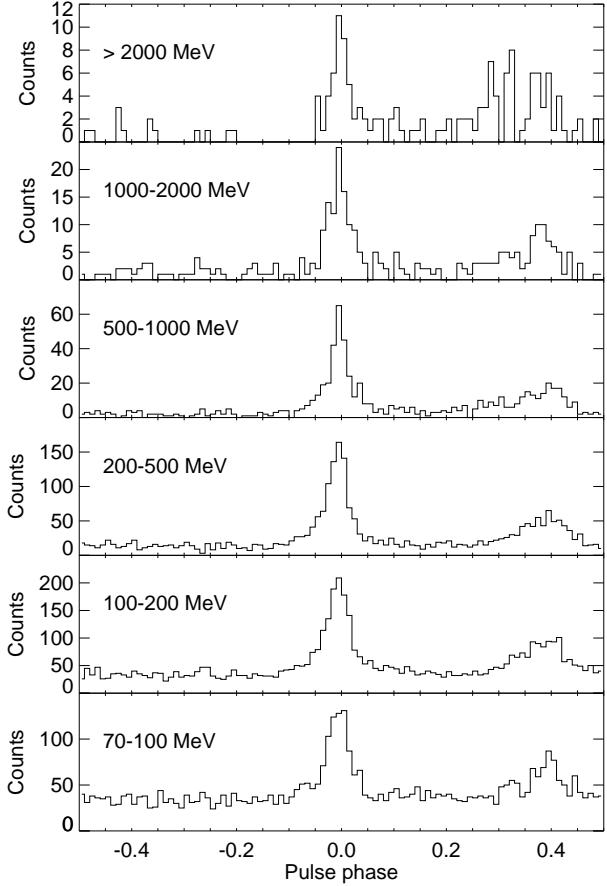}}
\caption{Crab pulsar phase histograms for various $\gamma$-ray
energy bands.  Zero phase is set by the radio ephemeris.}
\label{fig_phasehist} \end{figure}

Pulse phase histograms for several energy bands are shown in
Fig.~\ref{fig_phasehist}.  The main pulse peak, near phase
0.0, is the most appropriate feature for timing.  The main
peak is similar across the energy range from 70~MeV to 2~GeV
(\cite{fierro}). The peak width is about 0.05 in phase, and
appears somewhat narrower at high energies.  There is no
obvious shift of the peak centroid with energy.

\begin{figure}
\resizebox{\hsize}{!}{\includegraphics{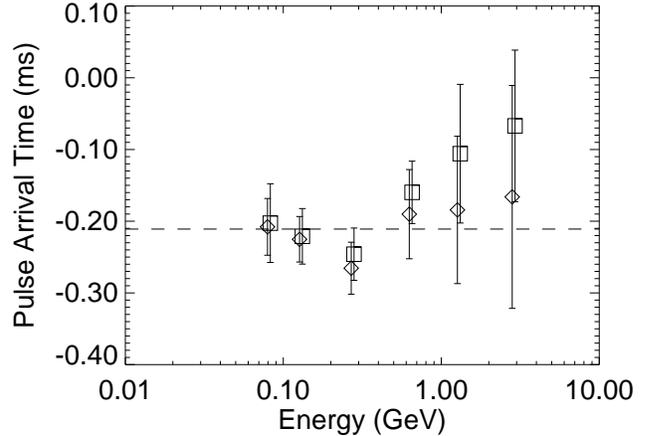}}
\caption{Pulse arrival time versus energy for the Crab.  The
diamonds indicate the average arrival time for photons within
the main pulse for each energy band.  The squares indicate
the centroid of a Lorentzian fit to the pulse profile for
each energy band.  The energies plotted are the median energy
for each band; the diamonds are shifted slightly in energy
for clarity.  The dashed line is the centroid of a Lorentzian
fit to the pulse profile for all energies above 70~MeV.}
\label{fig_centroid} \end{figure}

To study the energy dependence of the speed of light, we
measured the main peak pulse arrival time in each energy band.
We did this in two ways.  First, we calculated the average
arrival time for photons in the main peak. We found the
average time for each energy band using photons with phases
between -0.0464 and 0.0336, an interval centered on the mean
arrival time for all photons used in this analysis.  Second,
we parameterized the pulse arrival times by fitting a
Lorentzian to the pulse profile, within the same phase range
specified above, for each energy band.  Before fitting, a
constant rate equal to the average rate between phases -0.4
and -0.2 was subtracted.  The resultant was then fit with a
Lorentzian using a gradient-expansion algorithm to compute a
non-linear least squares fit.  The fits were all acceptable
with $\chi^2$ in the range 2.9 to 7.6 for 5 degrees of
freedom.

Fig.~\ref{fig_centroid} shows the pulse arrival times
calculated via both methods.  The errors in
Fig.~\ref{fig_centroid} correspond to $\Delta \chi^2 = 1$
(68\% confidence).  The energy of each point is the median
photon energy for each energy band. For the highest energy
band, the median energy is substantially lower than the
average, 2.9~GeV versus 5.0~GeV.  The zero pulse phase is set
by the radio ephemeris.  The pulse arrival time for all
photons used in this analysis is shown as a dashed line and
differs by 0.21~ms from the radio zero phase.  This is within
the error in the radio dispersion correction (\cite{nice}). 
We note that errors in the radio zero phase can broaden the
gamma-ray peak, but will not induce an energy dependent shift
in the gamma-ray pulse arrival time.  The accuracy of the
pulse arrival time determination for the Lorentzian fit is
0.07~ms ($\Delta \chi^2 = 3.84$ or 95\% confidence for a
single parameter of interest) in the 100--200~MeV band and
0.21~ms (95\% confidence) in the highest energy band.  The
accuracy in the highest energy band is limited mainly by
statistics.

It is apparent from the figure that there is no statistically
significant variation in pulse arrival time with energy. To
place an upper bound on any energy dependence in the speed of
light, we compare the arrival time for photons with energies
above 2~GeV (median energy 2.93~GeV) to that for the
70--100~MeV band (median energy 82.8~MeV).  The 95\%
confidence upper limit on the difference of the arrival times
is 0.35~ms.  Adopting a distance to the Crab of 2.2~kpc
(\cite{zombeck}), this leads to a lower limit on the energy
scale of quantum gravity effects on the speed of light of
$E_{QG} > 1.8 \times 10^{15} \, \rm GeV$ (95\% confidence).
This limit lies below the range of interest, but within an
order of magnitude of some predictions in the context of
string theory (\cite{witten}).

\section{Discussion}

Other effects which could also produce an energy dependent
delay in photon arrival times include energy dependent
dispersion due to the strong gravitational field near the
neutron star, purely electromagnetic dispersion, an energy
dependence in the emission location, or an intrinsic energy
dependence in the emission time.  The effect of any energy
dependent dispersion due to the strong gravitational field
near the neutron star is likely to be small because, even if
emitted from the neutron star surface, photons traverse the
region of high gravitational fields within about 0.1~ms. 
Allowing a fractional change in the speed of light equal to
the dimensionless field strength at the neutron star surface,
$GM/Rc^{2} \approx 0.2$, where $M \approx 1.4 \, M_{\odot}$ is
the neutron star mass and $R \approx \rm 10 \, km$  is the
neutron star radius, the difference in arrival times would be
only 0.02~ms.  The actual energy dependent change in the speed
of light is likely to be much smaller than 0.2.  Any
significant purely electromagnetic dispersion at MeV energies
and above can be excluded based on the dispersions measured at
lower energies.

An energy dependence in the photon emission location or
intrinsic emission time could produce a significant energy
dependent time delay.  While the possibility that precise
tuning of the emission locations or times for various energy
photons could cancel an energy dependent dispersion arising
from quantum gravity effects, we consider such a coincidence
unlikely, although not excluded, and interpret our lack of
detection of any energy dependence in arrival times as
constraining both the energy dependent dispersion and the
emission location and time.  In this case, the average
emission location, projected along our line of sight, for
photons at energies in the 70-100 MeV band must lie within
110~km of that for photons above 2~GeV, within 50~km of that
for 0.5--1.0~GeV photons, and within 150~km of that for radio
photons. 

It is encouraging that the analysis shows that it is possible
to time the Crab pulsar at gamma-ray energies to an accuracy
of 0.07~ms (95\% confidence) given adequate statistics. 
Detection of pulsations from the Crab at 50--100~GeV could
improve the limit on $E_{QG}$ by two orders of magnitude.  The
key question is whether the pulsations of the Crab and other
gamma-ray pulsars continue to such high energies. 
Observations of the Crab near 1~TeV show only unpulsed
emission (\cite{vacanti}) and the cutoff energy of the pulsed
emission is unknown.  If the Crab does pulse at 50--100~GeV,
detection of the pulses may be  possible in the near term with
low energy threshold atmospheric Cherenkov telescopes (ACTs),
such as STACEE (\cite{stacee}) and CELESTE (\cite{celeste}),
or in the longer term with a space-borne gamma-ray detector
such as GLAST (\cite{glast}).  The Crab pulsed signal may
extend only to the lowest energies accessible with the ACTs. 
Thus, measurement of a timing difference between two energy
bands might require contemporaneous measurements at other
wavelengths.  Both optical (\cite{smith}) and x-ray timing
(\cite{rots}) can exceed the accuracy of gamma-ray timing. 
However, the emission location for x-ray and optical photons
may differ from that of gamma-ray photons. If quantum gravity
does produce a first order correction to the dispersion
relation for electromagnetic waves, then measurement of the
pulse arrival time of the Crab at 50~GeV with an accuracy of
0.1~ms could be used to place a lower bound on $E_{QG} > 1.1
\times 10^{17} \, \rm GeV$. This is within the range, $10^{16}
- 10^{18} \rm \, GeV$, for the energy scale for quantum
gravity effects preferred in string theory (\cite{witten}).

If future measurements do reveal an energy dependence in
pulsar photon arrival times, then it will be difficult to
distinguish an energy dependent dispersion from an intrinsic
energy dependence in the emission location or emission time. 
This problem is common to all of the suggested astronomical
tests of quantum gravity effects.  Convincing proof for
quantum gravity effects will likely require detection of
energy dependent time delays in at least two different classes
of objects, preferably at vastly difference distances, i.e.
pulsars versus AGN or gamma-ray bursts, with all of the
detections compatible with the same value of $E_{QG}$.

\begin{acknowledgements}

I thank Paul Mende for useful discussions.  I acknowledge
partial support from NASA grant NAG5-7389.

\end{acknowledgements}

\end{document}